\documentclass[useAMS,usenatbib]{mn2e}

\usepackage{times}

\usepackage{graphicx}

\begin{document}

\def\X{\mathrm{x}}
\def\Y{\mathrm{y}}

\def\e{\mathrm{e}}
\def\b{\mathrm{b}}

\def\n{\mathrm{n}}
\def\p{\mathrm{p}}
\def\B{\mathrm{B}}
\def\eps{\varepsilon}
\def\mut{\widetilde{\mu}}
\def\pat{\sigma_p}

\def\d{\delta}

\def\mun{{\mu_\n}}
\def\mup{{\mu_\p}}

\def\sint{{\rm sin} \theta}
\def\cost{{\rm cos} \theta}
\def\cott{{\rm cot} \theta}
\def\tant{{\rm tan} \theta}

\def\E{{\mathcal{E}}}
\def\mut{\widetilde{\mu}}
\def\eps{\varepsilon}

\def\beq{\begin{equation}}
\def\eeq{\end{equation}}

\def\wiu{w^i_{\Y\X}}
\def\wid{w_i^{\Y\X}}

\title{Lagrangian perturbation theory of nonrelativistic  
rotating superfluid stars}

\author[N. Andersson, G.L. Comer and K. Grosart]
{N. Andersson$^1$, G.L. Comer$^2$ and K. Grosart$^1$\\
$^1$ School of Mathematics, University of Southampton, 
Southampton, SO17 1BJ, United Kingdom\\
$^2$ Department of Physics, Saint 
Louis University, St Louis, MO 63156,  USA}

\maketitle

\begin{abstract}
We develop a Lagrangian perturbation framework for rotating 
non-relativistic superfluid neutron stars. 
This leads to the first generalization of classic work 
on the stability properties of rotating stars to models
which account for the presence of potentially weakly coupled 
superfluid components. Our analysis is based on the standard
two-fluid model expected to be relevant for the conditions that 
prevail in the outer core of mature neutron stars.  
We discuss the implications of our results for dynamical
and secular instabilities of a simple neutron star 
model in which the two fluids are allowed to assume 
different (uniform) rotation rates. 
\end{abstract}

\maketitle


\section{Introduction}

Investigations into the stability properties of rotating self-gravitating
bodies are of obvious relevance to astrophysics. By improving our 
understanding of the relevant issues we can hope to shed light on 
the nature of the various dynamical and secular instabilities that may 
govern the spin-evolution of rotating stars.  The relevance of such 
knowledge for neutron star astrophysics may be highly significant. It is 
for example not clear (i) whether instabilities may affect nascent 
neutron stars leading to the slow birth spin rates inferred for radio 
pulsars, (ii) to what extent gravitational-wave driven 
instabilities may provide the explanation for the observed clustering of 
spin periods in Low-Mass X-ray Binaries, and (iii) if compact 
stars are likely to evolve through phases where  unstable oscillations 
lead to the emission of a detectable gravitational-wave signal (see the review article
by \citet{Anders03}
for a discussion of these issues and further references).

The aim of this (and a subsequent) paper is to 
develop a Lagrangian perturbation framework for rotating 
non-relativistic superfluid neutron stars. 
This leads to the first generalization of 
Friedman and Schutz's classic work \citep{friedman78:_lagran,friedman78:_secul_instab}
on the stability properties of rotating stars to models
which account for the presence of potentially weakly coupled 
superfluid components. We believe that our extension of 
the, by now 25 years old, single fluid results corresponds to a
significant step forwards for this area of research. 

Our analysis is based on the standard two-fluid model 
\citep{carter89:_covar_theor_conduc,comer94:_hamil_sf,
carter98:_relat_supercond_superfl,langlois98:_differ_rotat_superfl_ns,
lindblom00:_r_modes_superfl_neutr_stars,
sedrakian00:_perturb_sf_normal_mixtures,andersson01:_dyn_superfl_ns,
comer02:_zero_freq_subspace,lee03:_superfl_r_modes,
yoshida03:_inertial_sf,yoshida03:_r-modes_relat_superfl,
prix02:_variat_I,carter02:_canon_newton,carter03:_canon_newton}
expected to be relevant for the conditions that 
prevail in the outer core of mature neutron stars.  
In order to keep the analysis tractable we are forced to 
make many simplifying assumptions, but our model nevertheless 
includes the extra degrees of freedom associated with 
any two-fluid configuration. 
Most importantly, we allow our stationary background configurations to 
be such that the two fluids rotate at different 
rates. This is a key addition to the standard picture. It is  
a necessary inclusion since all seemingly viable models of the large 
Vela pulsar glitches are based on the notion of angular momentum transfer 
between different fluid components rotating at different rates. 
It is also an exciting addition to the standard model, because the 
additional degree of freedom may lead to  new phenomena.
An interesting example of this is the  recently discovered  superfluid 
two-stream instability \citep{acp03:_twostream_prl,prix04:_inertial,
andersson02:_twostream}. 

The layout of this paper is as follows: 
We begin by introducing the two-fluid model for superfluid stars
and the Lagrangian perturbation theory that we will employ
(Section~\ref{ingre}). We then revisit (in Section~\ref{sinflu}) what should 
be well-known results for a single (barotropic) fluid model.  We also 
revise those results in such a way that the main concepts will remain 
relevant for the two-fluid problem. Yet they are easier to introduce in 
the single fluid case. Having developed the necessary tools we then move 
on to the two-fluid case (Section~\ref{sfprob}). As we will see, the 
extension of 
the single fluid formulae is relatively straightforward provided that we 
only allow for the direct chemical coupling between the two fluids
[for the moment neglecting the entrainment effect 
\citep{andreev75:_three_velocity_hydro,
borumand96:_superfl_neutr_star_matter,comer03:_rel_ent,
comer04:_slow_rot_ent}, the role of 
which will be the main focus of a subsequent paper]. 
Section~\ref{instab} then provides a 
discussion of some implications of our results, which is followed in 
Section~\ref{rinstab} by specific implications for the r-mode 
instability and gravitational wave emission.  Finally, 
in Section~\ref{conclude} we briefly look ahead to the work that will be 
presented in a subsequent paper.    

\section{Two key ingredients} \label{ingre}

We begin our  discussion by introducing the 
two main ingredients of our analysis: The two-fluid model 
for neutron stars, and the Lagrangian perturbation theory 
developed by \cite{friedman78:_lagran,friedman78:_secul_instab}.  

\subsection{The two-fluid model for superfluid stars}

A mature neutron star is likely to contain several superfluid 
components. One would certainly expect superfluid neutrons to 
coexist with the crystal lattice of nuclei that makes
up the outer kilometer or so of the star. In the outer parts
of the fluid core superfluid neutrons are expected to coexist
with superconducting protons, while the deep core may contain
exotic states of matter like superfluid hyperons and perhaps 
even colour-superconducting deconfined quarks. The modelling
of a star comprising all these components is far beyond our current 
means. In fact, the parameters governing the possible states 
are uncertain to say the least. Given this, most investigations 
into the dynamics of superfluid neutron stars have considered 
a much simplified scenario that accounts for the presence
of two distinct fluid components. 

The two-fluid picture of superfluid neutron stars is based on the notion
that the outer core is dominated by superfluid neutrons, 
superconducting protons and electrons. Since the charged components
couple electromagnetically on a short timescale
they are assumed to move together. This assumption should be valid
as long as we consider dynamics that takes place on a timescale long 
compared to that associated with the electromagnetic interaction.
The outcome is a model describing the motion of the neutrons 
and the ``protons'' (a conglomerate of the charged components).

The equations that describe the two-fluid model 
are derived from an energy functional $E(n_\n,n_\p,w^2)$
where  $n_\X$ are the two number densities. Given the 
two fluid velocities $v^\X_i$ the quantity
$\wid \equiv v^\Y_i - v^\X_i $ ($\wiu \equiv v_\Y^i - v_\X^i $) represents 
the relative velocity between the two fluids, 
for ease of notation we define $w^2 \equiv \wiu \wid$. 
Throughout this paper
we will use the constituent indices $\X=\n,\p$ and $\Y\neq \X$.
This allows us to condense the various equations significantly, but in 
order to avoid confusion we should stress that repeated constituent indices
{\em never} imply summation, while repeated vector component indices 
{\em always} do. 
It is also worth pointing out that all calculations will be carried out 
in a coordinate basis. Hence, we will distinguish between co- and 
contravariant vectors etcetera.

Once we are supplied with the energy functional (the equation of state)
a straightforward variation yields
\begin{equation}
dE = \sum_{\X=\n,\p} \mu_\X dn_\X + \alpha dw^2 \ .
\end{equation}
This leads to the definition of the two chemical potentials
\begin{equation}
\mu_\X = \left( {\partial E \over \partial n_\X}
\right) _{n_\Y,w^2} \ .
\end{equation}
For later convenience, we introduce the notation
$\tilde{\mu}_\X=\mu_\X/m_\B$, where we take the two nucleon masses to be 
equal such that $m_\B=m_\n=m_\p$. We also see that the 
entrainment coefficient $\alpha$ follows from
\begin{equation}
\alpha = \left({\partial E \over \partial w^2}\right)_{n_\n, n_\p} \ .
\end{equation}

The dynamical equations that govern the two coupled fluids can be
derived either from a Newtonian variational principle 
\citep{prix02:_variat_I}
or from the Newtonian limit \citep{andersson01:_dyn_superfl_ns} of the 
fully relativistic 
equations derived by Carter and collaborators 
\citep{carter89:_covar_theor_conduc,comer94:_hamil_sf,
carter98:_relat_supercond_superfl,langlois98:_differ_rotat_superfl_ns}. 
The end result is two continuity equations
\begin{equation}
\partial_t n_\X + \nabla_i (n_\X v_\X^i) = 0 
\end{equation}
and two coupled Euler equations
\begin{eqnarray}
(\partial_t + v_\X^j \nabla_j ) \left(v^\X_i + \eps_\X \wid \right) 
  + \nabla_i \left(\Phi + \mut_\X \right) \hfill \nonumber \\
 +  \eps_\X
  w_j^{\Y\X} \nabla_i v_\X^j = 0 \ . 
\end{eqnarray}
Here we have defined
$\eps_\X= 2 \alpha/n_\X$. We also have the standard Poisson 
equation for the gravitational potential $\Phi$;
\beq
    \nabla^2 \Phi =
4 \pi m_\B G \left(n_\n + n_\p \right) \ . \label{poisson}
\eeq

We assume the background to be stationary and axisymmetric, with the two
fluids rotating around the $z$-axis with rates $\Omega_\n$ and
$\Omega_\p$ respectively. Hence we have
\begin{equation}
v^i_\X = \Omega_\X  \varphi^i \quad \textrm{and}\quad
\wiu = (\Omega_\Y - \Omega_\X) \varphi^i 
\end{equation}
with $\varphi^i$  given by
\begin{equation}
\varphi^i \partial_i = \partial_\varphi \ .
\end{equation}
In spherical coordinates, i.e. $ x^i \in \{r,\theta,\varphi \} $ , 
this vector has
the components $\varphi^i=(0,\,0,\,1)$, and its norm is  
$\varphi^i\varphi_i=r^2\sin^2\theta$.

In order to simplify the analysis, 
we restrict our attention to models with uniform rotation, i.e.
take $\Omega_\X$ to be constants.
Furthermore, in order to elucidate the details we will only consider the 
case of vanishing entrainment. That is, we let $\alpha = 0$. 
The motivation for this is simple: If we could not make progress 
even in this restricted case, it would be pointless to consider the 
much more complicated problem which includes the entrainment. 
However, as we will see, the case of vanishing entrainment works
out very neatly. Hence there is every reason for optimism, 
and we will return to the case $\alpha \neq 0$ in a subsequent paper.

\subsection{Lagrangian perturbation theory}

Our aim is to derive conserved quantities for the 
perturbations of the system of equations described above. By doing this
we hope to be able to provide criteria that can be used to 
decide when the oscillations of a rotating superfluid
neutron star are unstable. The need for such criteria is clear
given that i) the astrophysical relevance of the gravitational-wave 
driven r-mode instability may to a large extent depend on whether the 
star contains superfluid components \citep{Anders03}, and  
ii) the dynamical two-stream instability may set in above a 
critical relative rotation rate \citep{acp03:_twostream_prl,prix04:_inertial,
andersson02:_twostream}. A detailed analysis of both 
these problems clearly requires an improved understanding of 
the stability properties of superfluid stars.   

Following \cite{friedman78:_lagran,friedman78:_secul_instab}, we  analyze the problem within 
the Lagrangian perturbation formalism. The Lagrangian variation 
$\Delta Q$ of a 
quantity $Q$ is related to the Eulerian variation $\delta Q$  
by
\beq
\Delta Q = \delta Q + \pounds_\xi Q
\eeq
where the Lie derivative $\pounds_\xi$ has the meaning
\beq
\pounds_\xi f = \xi^i \nabla_i f 
\eeq
for scalars, 
\beq
\pounds_\xi v^i = \xi^j\nabla_j v^i - v^j \nabla_j \xi^i 
\eeq
for contravariant vectors, and
\beq
\pounds_\xi v_i = \xi^j\nabla_j v_i + v_j \nabla_i \xi^j 
\eeq
for covariant vectors. 

The Lagrangian change in the fluid velocity follows from
\beq
\Delta v^i = \partial_t \xi^i
\eeq
where $\xi^i$ is the Lagrangian displacement.
Given this, and
\beq
\Delta g_{ij} = \nabla_i \xi_j + \nabla_j \xi_i
\eeq
we have
\beq
\Delta v_i = \partial_t \xi_i + v^j\nabla_i \xi_j + v^j \nabla_j \xi_i 
\ .
\eeq

It is also useful to note that the Eulerian variations are given by
\beq
\delta v^i = \partial_t \xi^i + v^j \nabla_j \xi^i - \xi^j \nabla_j v^i
\label{vup}\eeq
and
\beq
\delta v_i = \partial_t \xi_i + v^j \nabla_j \xi_i - \xi^j \nabla_j v_i
\label{vdn}\eeq
(quite obviously, since $v_i = g_{ij}v^j$ and $\nabla_k g_{ij}=0$).

\section{Revisiting and revising the single fluid problem} \label{sinflu}

In order to lay the foundation for our analysis of the superfluid 
problem, it is useful 
to revisit the analysis of an ordinary perfect fluid star. In 
doing this we want to stay as close as possible to the equations used to 
describe the two-fluid problem. We know from previous work that the 
relevant model to compare to is a barotropic perfect fluid, eg.~since 
the g-modes are absent from the pulsation spectrum of a non-rotating 
superfluid model 
\citep{lee95:_nonrad_osc_superfl_ns,andersson01:_dyn_superfl_ns,
comer02:_zero_freq_subspace}. 
Furthermore, we prefer to work 
with the number density $n$, the fluid velocity $v_i$ and the chemical 
potential $\mu$ rather than the pressure $P$. 

\subsection{The perturbation equations}

From thermodynamic principles 
we know that, for a barotropic ordinary fluid we have $E=E(n)$, 
and 
\beq
d \mu = {1 \over n } d P \ .
\eeq
This allows us to write the standard Euler equation as 
\beq
(\partial_t + v^j \nabla_j) v_i +  \nabla_i (\tilde{\mu} + \Phi) = 0
\label{euleq1}\eeq
where  $\tilde{\mu} = \mu/m_\B$ as before. In addition we have the 
continuity equation
\beq
\partial_t n + \nabla_i (n v^i) = 0
\eeq
and the Poisson equation for the gravitational potential
\beq
\nabla^2 \Phi = 4\pi m_\B G n \ .
\eeq

We now want to perturb these equations. First of all,  
conservation of mass for the perturbations is readily expressed as
\beq
\Delta n = - n \nabla_i \xi^i
\longrightarrow 
\delta n = - \nabla_i (n  \xi^i) \ .
\eeq
Consequently, the perturbed gravitational potential follows from 
\beq
\nabla^2 \delta \Phi = 4 \pi G m_\B \delta n = - 4\pi G m_\B \nabla_i(n  
 \xi^i) \ .
\eeq

In order to perturb the Euler equations we first rewrite 
Eq.~(\ref{euleq1}) as
\beq
(\partial_t +\pounds_v) v_i +  \nabla_i \left( \tilde{\mu}  + \Phi
- { 1 \over 2}  v^2 \right) = 0 \ .
\label{euleq2}
\eeq
This form is particularly useful since the Lagrangian variation 
commutes with the operator $\partial_t + \pounds_v$, i.e. 
\beq
\Delta (\partial_t+ \pounds_v) v_i =  (\partial_t + \pounds_v) \Delta v_i
\ .
\eeq
Perturbing (\ref{euleq2}) we thus have
\beq
(\partial_t +\pounds_v) \Delta v_i + \nabla_i \left(  \Delta \tilde{\mu} 
+ \Delta \Phi
- { 1 \over 2} \Delta( v^2) \right) = 0 \ .
\label{peuls}\eeq

We want to rewrite this equation in terms of the displacement vector 
$\xi$. After some algebra one finds that 
\begin{eqnarray}
\partial_t^2 \xi_i + 
2 v^j \nabla_j \partial_t \xi_i +  (v^j \nabla_j)^2 \xi_i + 
\nabla_i \delta \Phi + 
\xi^j \nabla_i \nabla_j \Phi 
\nonumber \\
 -   (\nabla_i \xi^j) \nabla_j\tilde{\mu}
+\nabla_i \Delta \tilde{\mu} 
  = 0 \ .
\label{peul2}\end{eqnarray}
Finally, we need
\begin{eqnarray}
\Delta \tilde{\mu} = \delta \tilde{\mu} + \xi^i\nabla_i \tilde{\mu} = 
\left( {\partial \tilde{\mu} \over \partial n} \right) \delta n 
+  \xi^i\nabla_i \tilde{\mu} \nonumber \\ 
= \quad - \left( {\partial \tilde{\mu} \over \partial n} \right) \nabla_i 
(n\xi^i) +  \xi^i\nabla_i \tilde{\mu} \ .
\end{eqnarray}
Given this, we have arrived at the following form for the perturbed
Euler equation
\begin{eqnarray}
\partial_t^2 \xi_i + 
2  v^j \nabla_j \partial_t \xi_i + (v^j \nabla_j)^2 \xi_i + 
 \nabla_i \delta \Phi 
+  \xi^j \nabla_i \nabla_j ( \Phi + \tilde{\mu}) \nonumber \\
- \nabla_i \left[ \left( {\partial \tilde{\mu} \over \partial n} \right) 
\nabla_j (n\xi^j) 
\right]
  = 0 \ . \label{peul3}
\end{eqnarray}
This equation should be compared to Eq.~(15) of 
\cite{friedman78:_lagran}.

\subsection{Conserved quantities: The canonical energy/angular momentum}

Having derived the perturbed Euler equations, we want to construct 
conserved quantities that can be used to assess the stability of the 
system.  To do this, we first multiply Eq.~(\ref{peul3}) by the number 
density $n$, and then write the result (schematically) as
\beq
A \partial_t^2 \xi + B \partial_t \xi + C \xi  = 0 \ .
\eeq
We are omitting the indices, since there should be little risk of 
confusion. 

Defining the standard inner product
\beq
\left< \eta^i,\xi_i \right> = \int \eta^{i*} \xi_i dV
\eeq
where the asterisk denotes complex conjugation, one can readily show that
\beq
\left< \eta, A\xi \right>=\left< \xi,A\eta \right>^* \qquad \mbox{ and } 
\qquad
\left< \eta,B\xi \right> = - \left< \xi,B\eta \right>^* \ .
\eeq
The latter requires the background relation $\nabla_i (n v^i) = 0$, 
and holds provided that $n \to 0$ at the surface of the star. A slightly 
more involved calculation leads to 
\beq
\left< \eta, C\xi \right> = \left< \xi, C\eta \right>^* \ .
\eeq
In particular, we use 
\beq
\int n \eta^{i*} \nabla_i \left[ \left( {\partial \tilde{\mu} \over 
\partial n} \right) \nabla_j (n\xi^j)  \right] dV  = 
\int n \xi^i \nabla_i \left[ \left( {\partial \tilde{\mu} \over \partial 
n} \right) \nabla_j (n\eta^{j*})  \right] dV 
\eeq
which (again) holds as long as  $n\to 0$ at the surface, and
\begin{eqnarray}
\int n \eta^{i*} \nabla_i \delta_\xi \Phi dV =
{ 1 \over 4 \pi m_\B G} \int  g^{ij} \delta_\xi \Phi \nabla_i \nabla_j \delta_\eta 
\Phi^* dV  \nonumber \\
= - { 1 \over 4 \pi m_\B G} \int g^{ij} (\nabla_i \delta_\xi \Phi) (\nabla_j \delta_\eta 
\Phi^*) dV  \nonumber \\
= \int n \xi^j \nabla_j (\delta_\eta \Phi)^* dV  \ . 
\end{eqnarray}

Inspired by the fact that the momentum conjugate to
$\xi^i$ is $\rho(\partial_t + v^j \nabla_j)\xi^i$, we now consider the
symplectic structure
\beq
W(\eta,\xi) = \left<\eta, A\partial_t \xi + { 1 \over 2}B\xi\right>
- \left< A\partial_t \eta + { 1 \over 2}B \eta, \xi\right>
\label{Wdef}
\eeq 
where $\eta$ and $\xi$ both solve the perturbed Euler equation. 
Given this, it is straightforward to show that $W(\eta,\xi)$ is 
conserved, i.e.
$\partial_t W = 0$. This leads us to define the \emph{canonical energy} 
of the system as
\beq
E_c = { m_\B \over 2} W (\partial_t \xi,\xi) = { m_\B \over 2} 
\left\{ \left< \partial_t \xi , A \partial_t \xi
\right> 
 + \left< \xi, C \xi \right> \right\} \ .
\eeq
After some manipulations, we arrive at the following explicit expression 
\begin{eqnarray}
E_c = { 1 \over 2} \int \left\{ 
\rho |\partial_t \xi|^2 
- \rho | v^j \nabla_j \xi_i|^2
+ \rho \xi^i \xi^{j*}\nabla_i \nabla_j (\tilde{\mu} + \Phi) \right. \nonumber \\
+\left.  
\left( {\partial \mu \over \partial n} \right) | \delta n |^2
-   { 1 \over 4\pi G} |\nabla_i \delta \Phi|^2
\right\} dV 
\end{eqnarray}
which can be compared to Eq.~(45) of \cite{friedman78:_lagran}.

In the case of an axisymmetric system, eg. a rotating star, we can also 
define a \emph{canonical angular momentum} as 
\beq
J_c = - { m_\B \over 2} W (\partial_\varphi \xi, \xi) = -\mbox{ Re }  
\left<
\partial_\varphi \xi, A\partial_t \xi + { 1 \over 2} B\xi \right> \ .
\eeq
 The proof that this quantity is conserved relies on the fact that 
(i) $W(\eta, \xi)$ is conserved for any two solutions to the
perturbed Euler equations, and (ii) $\partial_\varphi$ commutes
with $\rho v^j \nabla_j$ in axisymmetry, which means that if
$\xi$ solves the Euler equations then so does $\partial_\varphi \xi$.

As elucidated by \cite{friedman78:_lagran,friedman78:_secul_instab}, the stability analysis is 
complicated by the presence of so-called ``trivial'' displacements.
These trivials can be thought of as ``integration constants'' 
representing a relabeling of the physical fluid elements. A trivial 
displacement $\zeta^i$ leaves the physical quantities unchanged, i.e.~is 
such that $\delta n = \delta v^i = 0$.
This means that we must have
\begin{eqnarray}
\nabla_i (\rho \zeta^i) &=& 0 \ , \\
  ( \partial_t + \pounds_v ) \zeta^i &=& 0 \ .
\end{eqnarray}
The solution to the first of these equations can be written
\beq
\rho \zeta^i = \epsilon^{ijk} \nabla_j \chi_k 
\eeq
where, in order to satisfy the second equations, the vector 
$\chi_k$ must have time-dependence such that
\beq
( \partial_t + \pounds_v) \chi_k = 0 \ .
\eeq
This means that the trivial displacement will remain constant along the 
background fluid trajectories. Or, as \cite{friedman78:_lagran} put it, 
the ``initial relabeling is carried along with the unperturbed motion.''

The trivials may cause trouble because they affect the canonical 
energy. Before one can use the canonical energy to 
assess the stability of a rotating configuration one must deal with this
``gauge problem.'' The way to do this is to ensure that the 
displacement vector $\xi$ is orthogonal to all trivials. 
A prescription for doing this is provided by \cite{friedman78:_lagran}.
In particular, they show that the required  
canonical perturbations preserve the 
vorticity of the individual fluid elements. Most importantly,  
one can also prove that 
a normal mode solution is orthogonal to the trivials. 
Thus, normal mode solutions can serve as canonical initial 
data, and be used to assess stability.

\subsection{Example: Instabilities of rotating perfect fluid stars}
\label{pfinstab}

The importance of the canonical energy stems from the fact that 
it can be used to test the stability of the system.
In particular, we note that:
\begin{itemize}
\item Dynamical instabilities are only possible for motions
such that $E_c=0$. This makes intuitive sense
since the amplitude of 
a mode for which $E_c$ vanishes can grow 
without bounds and still obey the conservation laws.

\item If the system is coupled to radiation (eg. gravitational waves) 
which carries positive energy away from the system (which should be 
taken to mean that
$\partial_t E_c < 0$) then any initial data 
for which $E_c<0$ will lead to an unstable evolution.

\end{itemize}

Consider a real frequency normal-mode
solution to the perturbation equations, a solution of form
$\xi = \hat{\xi} e^{i(\omega t+m\varphi)}$. 
One can readily show that the associated canonical energy becomes
\beq
E_c = \omega \left[  \omega \left<{\xi} , A {\xi}\right> - { i \over 2} 
\left<{\xi} , B{\xi}\right> \right] \label{Ec}
\eeq
where the expression in the bracket is  real valued.
For the canonical angular momentum we get
\beq
J_c = -m \left[  \omega \left<{\xi} , A {\xi} \right> - { i \over 2} 
\left< {\xi} , B{\xi} \right> \right] \ .
\label{Jc}
\eeq 

Combining Eq.~(\ref{Ec}) and Eq.~(\ref{Jc}) we see that, for real 
frequency modes we will have
\beq
E_c = - {\omega \over m} J_c = \sigma_p J_c
\label{EJrel}
\eeq
where $\sigma_p$ is the pattern speed of the mode.

Now notice that Eq.~(\ref{Jc}) can be rewritten as
\beq
{ J_c \over \left< \hat{\xi}, \rho\hat{\xi} \right>} = - m\omega + 
m{ \left<{\xi}, i\rho \vec{v}\cdot \nabla {\xi} \right> 
\over  \left< \hat{\xi}, \rho\hat{\xi} \right>} \ .
\eeq
Using cylindrical coordinates, and $v^j = \Omega \varphi^j$, one can 
show that
\beq
-i\rho {{\xi}}_i^* v^j \nabla_j {\xi}^i = \rho \Omega [ m |\hat{\xi}|^2 
+ i ({\hat{\xi}}^* \times \hat{\xi})_z] \ .
\eeq
But
\beq
| ({\hat{\xi}}^* \times \hat{\xi})_z | \le | \hat{\xi} |^2
\eeq
and we must have (for uniform rotation)
\beq
\sigma_p - \Omega \left( 1 + {1\over m} \right) \le { J_c/m^2 \over 
\left<\hat{\xi}, \rho\hat{\xi}\right>}
\le \sigma_p - \Omega \left( 1 - {1\over m} \right) \ .
\label{ineq1}
\eeq

Eq.~(\ref{ineq1}) forms an integral part of the 
\cite{friedman78:_secul_instab} proof that
rotating perfect fluid stars are generically unstable in the presence of 
radiation. The argument is as follows: Consider modes with 
finite frequency in the $\Omega \to 0$ limit. Then Eq.~(\ref{ineq1}) 
implies that 
co-rotating modes (with $\sigma_p>0$) must have $J_c>0$, while 
counter-rotating modes
(for which $\sigma_p < 0$) will have $J_c<0$. In both cases $E_c>0$, 
which means 
that both classes of modes are stable. Now consider a small region near 
a point where $\sigma_p=0$ (at a finite rotation rate). Typically, this 
corresponds to a point where the initially counter-rotating mode becomes 
co-rotating. In this region 
$J_c<0$. However, $E_c$ will change sign at the point where $\sigma_p$
(or, equivalently, the frequency $\omega$) vanishes. Since the mode was 
stable
in the non-rotating limit this change of sign indicates the onset of 
instability at a critical rate of rotation. 

\subsection{Example: The r-mode instability}

In order to further demonstrate the usefulness of the canonical energy, 
let us prove the instability of the single-fluid r-modes. 

For a general inertial mode we have (cf. \citet{lockitch99:_r-modes} who provide a 
discussion of the single fluid problem using notation which closely resembles the one we adopt here) 
\beq
\vec{v} \sim \delta \vec{v} \sim \dot{\vec{\xi}} \sim \Omega \qquad
\mbox{ and } \qquad 
\delta \Phi \sim \delta n \sim \Omega^2 \ .
\eeq
If we also assume axial-led modes, like the r-modes, 
then we have $\delta v_r \sim \Omega^2$
and the continuity equation leads to 
\begin{equation}
\nabla \cdot \delta \vec{v} \sim \Omega^3 \rightarrow 
\nabla \cdot \vec{\xi} \sim \Omega^2 \ .
\end{equation}

Under these assumptions we find that $E_c$ becomes (to order $\Omega^2$)
\begin{equation}
E_c \approx {1\over 2} \int \rho \left[  |\partial_t {\xi}|^2
- |{v}\cdot\nabla{\xi}|^2  +  \xi^{i*} \xi^{j} \nabla_i\nabla_j (
\Phi + \tilde{\mu})
\right]  dV \ .
\label{ec1}
\end{equation}
We can rewrite the last term using the equation governing the 
axisymmetric equilibrium.
Keeping only terms of order $\Omega^2$ we have
\begin{equation}
\xi^{i*} \xi^{j} \nabla_i\nabla_j (
\Phi + \tilde{\mu}) 
\approx { 1 \over 2} \Omega^2  \xi^{i*} \xi^{j}
\nabla_i \nabla_j (r^2 \sin^2 \theta) \ .
\end{equation}
A bit more work then leads to
\begin{equation}
{ 1 \over 2} \Omega^2  \xi^{i*} \xi^{j}
\nabla_i \nabla_j (r^2 \sin^2 \theta) =  
\Omega^2 r^2 \left[ \cos^2 \theta |\xi^\theta|^2 + \sin^2\theta | 
\xi^\varphi|^2  \right]
\end{equation}
and
\begin{eqnarray}
| v^i  \nabla_i \xi_j |^2 = \Omega^2 \left\{ m^2 | \xi |^2  - 2imr^2 
\sin \theta \cos \theta \left[
\xi^\theta \xi^{\varphi *} - \xi^\varphi \xi^{\theta *} \right] \right. \nonumber \\
+ \left.  
r^2 \left[ \cos^2 \theta |\xi^\theta|^2 + \sin^2\theta | \xi^\varphi|^2  
\right] \right\} 
\end{eqnarray}
which means that the canonical energy can be written in the form
\begin{eqnarray}
E_c \approx - { 1 \over 2} \int \rho \left\{ 
(m \Omega -\omega)(m\Omega +\omega) | \xi |^2 \right. \qquad \qquad \nonumber \\ 
 \left. -  2im \Omega^2 r^2 \sin \theta \cos \theta \left[
\xi^\theta \xi^{\varphi *} - \xi^\varphi \xi^{\theta *} \right]
\right\} dV
\end{eqnarray}
for an axial-led mode.

Introducing the axial stream function,
\begin{eqnarray}
\xi^\theta &=& - { iU \over r^2 \sin \theta} \partial_\varphi Y_l^m 
e^{i\omega t} \ , \\
\xi^\varphi &=& { iU \over r^2 \sin \theta} \partial_\theta Y_l^m 
e^{i\omega t} \ , 
\end{eqnarray} 
where $Y_l^m=Y_l^m(\theta,\varphi)$ are the standard spherical harmonics,
we have
\begin{equation}
|\xi|^2 = { |U|^2 \over r^2} \left[  { 1 \over \sin^2 \theta} 
| \partial_\varphi Y_l^m |^2 + |\partial_\theta Y_l^m|^2 \right] 
\end{equation}
and
\begin{eqnarray}
ir^2 \sin \theta \cos \theta \left[
\xi^\theta \xi^{\varphi *} - \xi^\varphi \xi^{\theta *} \right] \qquad \qquad \qquad \qquad \nonumber \\
= { 1 \over r^2} { \cos \theta \over \sin \theta} m |U|^2 
\left[ Y_l^m \partial_\theta Y_l^{m*} +  Y_l^{m *} \partial_\theta 
Y_l^{m}\right] \ . 
\end{eqnarray}

After performing the angular integrals, we find that 
\begin{eqnarray}
E_c = - { l(l+1) \over 2} \left\{ (m\Omega -\omega)(m\Omega+\omega) -  
{2 m^2 \Omega^2 \over l(l+1)} \right\} \int \rho |U|^2 dr 
\end{eqnarray}
Combining this with the r-mode frequency \citep{lockitch99:_r-modes}
\begin{equation}
\omega = m\Omega \left[ 1 - { 2 \over l(l+1)} \right] 
\end{equation}
we see that  $E_c < 0$ for all $l>1$ r-modes, i.e.~they are all unstable.
The $l=m=1$ r-mode is a special case, leading to $E_c=0$. 

\section{The superfluid problem} \label{sfprob}

In this Section we generalise the Lagrangian perturbation analysis to 
the two-fluid model for superfluid neutron stars. In order to 
simplify matters we only consider the case of vanishing entrainment. 
The, significantly more complicated, general case will be discussed in a 
subsequent paper.

\subsection{The perturbation equations}

As in the single-fluid problem, we begin by deriving the 
equations governing Lagrangian perturbations of the system.
Assuming vanishing entrainment, i.e. letting $\alpha=0$, we have the Euler equations
\begin{eqnarray}
  \left( \partial_t   + \pounds_{v_\X}\right) v^\X_i   +  \nabla_i 
       \left(\Phi + \tilde{\mu}_\X - { 1\over 2} v_\X^2\right)
 = 0 
\label{eul1}\end{eqnarray}
where we recall that $\X=\n$ or $\p$.
Clearly, we must introduce two distinct
Lagrangian displacement vectors $\vec{\xi}_\X$. 
To distinguish between the two possibilities we use 
variations $\Delta_\X$ such that
\beq
\Delta_\X Q = \delta Q + \pounds_{\xi_\X} Q \ .
\eeq
The analogous generalisation of the single-fluid 
formulae for co- and contravariant vectors is straightforward. 

The two Lagrangian variations are naturally introduced
in such a way that
\beq
\Delta_\X v_\X^i = \partial_t \xi_\X^i
\eeq
which leads to
the perturbed continuity equations taking the form
\beq
\Delta_\X  n_\X = - n_\X \nabla_i \xi_\X ^i \longrightarrow
\delta n_\X = - \nabla_i (n_\X \xi_\X^i) \ .
\eeq

With these definitions, it is very easy to derive the perturbed 
Euler equations. Simply comparing Eq.~(\ref{euleq2}) to Eq.~(\ref{eul1}) 
we see that we must have
\begin{eqnarray}
\partial_t^2 \xi^\X_i + 
2 v_\X^j \nabla_j \partial_t \xi^\X_i +  (v_\X^j \nabla_j)^2 \xi^\X_i + 
\nabla_i \delta \Phi + 
\xi_\X^j \nabla_i \nabla_j \Phi \nonumber \\
- (\nabla_i \xi_\X^j) \nabla_j\tilde{\mu}_\X
+\nabla_i \Delta_\X \tilde{\mu}_\X 
  = 0 \ .
\label{sfpeul2}
\end{eqnarray}
To express this in terms of the displacement vectors we need
\begin{eqnarray}
\Delta_\X \tilde{\mu}_\X = 
\delta \tilde{\mu}_\X + \xi_\X^i\nabla_i \tilde{\mu}_\X \hspace{5.5cm} \nonumber \\ 
= \left( {\partial \tilde{\mu}_\X \over \partial n_\X} \right)_{n_\Y} 
\delta n_\X 
+ \left( {\partial \tilde{\mu}_\X \over \partial n_\Y} \right)_{n_\X} 
\delta n_\Y +  \xi_\X^i\nabla_i \tilde{\mu}_\X  \hspace{2cm} \nonumber \\
=
- \left( {\partial \tilde{\mu}_\X \over \partial n_\X} \right)_{n_\Y} 
\nabla_i (n_\X\xi_\X^i)
- \left( {\partial \tilde{\mu}_\X \over \partial n_\Y} \right)_{n_\X} 
\nabla_i (n_\Y \xi_\Y^i)
+  \xi_\X^i\nabla_i \tilde{\mu}_\X
\end{eqnarray}
and we arrive at the following form for the perturbed
Euler equations
\begin{eqnarray}
\partial_t^2 \xi^\X_i + 
2  v_\X^j \nabla_j \partial_t \xi^\X_i + (v_\X^j \nabla_j)^2 \xi^\X_i + 
 \nabla_i \delta \Phi 
+ \xi_\X ^j \nabla_i \nabla_j (\Phi + \tilde{\mu}_\X)
\nonumber \\
- \nabla_i \left[ \left( {\partial \tilde{\mu}_\X \over \partial n_\X}
 \right)_{n_\Y} \nabla_j (n_\X\xi_\X^j) + 
\left( {\partial \tilde{\mu}_\X \over \partial {n_\Y}} \right)_{n_\X} 
\nabla_j (
n_\Y\xi_\Y^j)
\right]
  = 0 \ .
\label{sfpeul3}
\end{eqnarray}

From this equation it is clear that  the two fluids are coupled. 
In order to proceed we need to understand the nature of this coupling 
better. 
In particular, we note that the perturbed 
gravitational potential depends on both displacement vectors.
We have
\beq
\nabla^2 \delta \Phi = 4\pi m_\B G ( \delta n_\n + \delta n_\p) = 
- 4\pi m_\B G \nabla_i (n_\n\xi_\n^i + n_\p \xi^i) \ .
\eeq
Since this is a linear equation we can write the solution as
\beq
\delta \Phi = \delta \Phi_\n + \delta \Phi_\p = \sum_{\X=\n,\p} 
\delta \Phi_\X
\eeq
where we define
\beq
\nabla^2 \delta \Phi_\X = 4\pi m_\B G  \delta n_\X  = - 4\pi m_\B G
\nabla_i (n_\X\xi_\X^i) \ .
\eeq
%

In analogy with the single fluid case, we can write the perturbed
Euler equations 
in the schematic form (after multiplying Eq.~(\ref{sfpeul3}) by $n_\X$) 
\beq
A_\X \partial_t^2 \xi_\X + B_\X \partial_t \xi_\X + C_\X \xi_\X + D_\X \xi_\Y = 0
\ .
\label{sfeom}\eeq
It should be noted that the first three terms are obvious 
generalisations 
of the single fluid case. Now they pertain to each of the two fluids. 
The last term is new, and describes the coupling between the 
fluids. Explicitly, it takes the form
\beq
D_\X \xi_\Y = -n_\X \nabla_i \left[\left( {\partial \tilde{\mu}_\X \over 
\partial
n_\Y} \right)_{n_\X} \nabla_j (n_\Y \xi_\Y^j) \right] + n_\X \nabla_i \delta 
\Phi_\Y \ .
\eeq
From this we see that the fluids are coupled (i) ``chemically'' through 
the equation of state, and (ii) ``gravitationally'' because of the fact 
that variations in one of the number densities affects the gravitational 
potential, which then influences the other fluid. 

\subsection{Conserved quantities in the superfluid case}

We want to derive conserved quantities similar to those in the 
single-fluid case. 
To do this we again use the inner product. Given the results
from Sec.~\ref{sinflu}, it is easy to show that we have the following 
symmetries 
\begin{eqnarray}
\left<\eta_\X,A_\X \xi_\X\right> &=& \left<\xi_\X,A_\X \eta_\X\right>^*   \ , \label{sfsymm1}\\
\left<\eta_\X,B_\X \xi_\X\right> &=& - \left<\xi_\X,B_\X \eta_\X\right>^* \ , \label{sfsymm2} \\
\left<\eta_\X,C_\X \xi_\X\right> &=& \left<\xi_\X,C_\X \eta_\X\right>^* \ , \label{sfsymm3}
\end{eqnarray}
where $\eta^i_\X$ can (at this point) be any vector field. 

Next we want to  introduce  symplectic structures that would be natural 
generalisations of the one we used to construct the canonical 
energy and angular momentum in the single fluid problem. 
To do this we consider two sets of solutions 
$[\xi_\n,\xi_\p]$ and $[\eta_\n, \eta_\p]$ to our perturbation \
equations. Then we define
\begin{eqnarray}
W_\n(\eta_n,\xi_n) 
=  \left< \eta_\n, A_\n \partial_t \xi_\n + 
\frac{1}{2} B_\n \xi_\n \right> \qquad \qquad \nonumber \\
 -  \left< A_\n \partial_t \eta_\n + \frac{1}{2} B_\n \eta_\n, \xi_\n 
\right> \ . 
\end{eqnarray}
Given this definition 
and the above symmetry relations, it is straightforward to show that
\beq
\partial_t W_\n = - \left<\eta_\n,D_\n \xi_\p\right> + \left<D_\n 
\eta_\p, \xi_\n\right> \neq 0 \ .
\eeq
Analogously we introduce
\begin{eqnarray}
W_\p (\eta_\p,\xi_\p) = \left< \eta_\p,A_\p \partial_t \xi_\p + 
\frac{1}{2} B_\p \xi_\p \right> \qquad \qquad \nonumber \\
-  \left< A_\p \partial_t \eta_\p + \frac{1}{2} B_\p \eta_\p, \xi_\p 
\right>  
\end{eqnarray}
which leads to 
\beq
\partial_t W_\p = -\left<\eta_\p,D_\p \xi_\n\right> + \left<D_\p 
\eta_\n, \xi_\p\right> \neq 0 \ .
\eeq

Intuitively one would expect the sum $W_\n+W_\p$ to be conserved. That 
is, the coupling terms in Eq.~(\ref{sfpeul3}) should facilitate 
non-dissipative energy transfer between the two fluids. We will now 
prove that this is, indeed, the case. 

Explicitly we have
\begin{eqnarray}
\partial_t W_\n = \int n_\n \eta_\n^{i*} \nabla_i \left[ 
\left( {\partial \tilde{\mu}_\n \over \partial n_\p} \right)_{n_\n} 
\nabla_j (n_\p\xi_\p^j) 
 \right] dV \hspace{1cm} \nonumber \\ - \int n_\n \xi_\n^i \nabla_i   \left[ 
\left( {\partial \tilde{\mu}_\n \over \partial n_\p} \right)_{n_\n} 
\nabla_j (n_\p \eta_\p^{j*})
\right] dV \qquad \nonumber \\
- \int n_\n \eta_\n^{i*} \nabla_i \delta_{\xi_\p} \Phi_\p dV 
+ \int n_\n \xi_\n^i \nabla_i (\delta_{\eta_\p} \Phi_\p)^* dV 
\label{dWn}\end{eqnarray}
and
\begin{eqnarray}
\partial_t W_\p = \int n_\p \eta_\p^{i*} \nabla_i \left[ 
\left( {\partial \tilde{\mu}_\p \over \partial n_\n} \right)_{n_\p} 
\nabla_j (n_\n\xi_\n^j) 
 \right] dV 
\hspace{1cm} \nonumber \\
- \int n_\p \xi_\p^i \nabla_i   \left[ 
\left( {\partial \tilde{\mu}_\p \over \partial n_\n} \right)_{n_\p} 
\nabla_j (n_\n \eta_\p^{j*})
\right] dV \qquad \nonumber \\
- \int n_\p \eta_\p^{i*} \nabla_i \delta_{\xi_\n} \Phi_\n dV 
+ \int n_\p \xi_\p^i \nabla_i (\delta_{\eta_\n} \Phi_\n)^* dV \ .
\label{dWp}
\end{eqnarray}
Consider the first two terms of (\ref{dWp}). Using the 
fact that 
\beq
\left( {\partial \tilde{\mu}_\p \over \partial n_\n} \right)_{n_\p}  = 
\left( {\partial \tilde{\mu}_\n \over \partial n_\p} \right)_{n_\n} 
\eeq 
and assuming that $n_\n$ and $n_\p$ both vanish at the surface\footnote{Somewhat artificially, 
we assume that the rotating background is such that the two fluids have a common surface.
In reality, the outer layers of a neutron star will not be superfluid and one would have to add
a single fluid envelope to our model. The analysis of such composite models is beyond the scope
of the present analysis.}
\begin{eqnarray}
\int n_\p \eta_\p^{i*} \nabla_i \left[ 
\left( {\partial \tilde{\mu}_\p \over \partial n_\n} \right)_{n_\p} 
\nabla_j (n_\n\xi_\n^j) 
 \right] dV \hspace{2cm} \nonumber \\
 = \int n_\n \xi_\n^i \nabla_i   \left[ 
\left( {\partial \tilde{\mu}_\n \over \partial n_\p} \right)_{n_\n} 
\nabla_j (n_\p \eta_\p^{j*})\right] dV  
\end{eqnarray}
and 
\begin{eqnarray}
\int n_\p \xi_\p^i \nabla_i   \left[ 
\left( {\partial \tilde{\mu}_\p \over \partial n_\n} \right)_{n_\p} 
\nabla_j (n_\n \eta_\n^{j*})
\right] dV \hspace{2cm} \nonumber \\
=  
\int n_\n \eta_\n^{i*} \nabla_i \left[ 
\left( {\partial \tilde{\mu}_\n \over \partial n_\p} \right)_{n_\n} 
\nabla_j (n_\p\xi_\p^j) 
 \right] dV  \ .
\end{eqnarray}
This means that, when the two expressions Eq.~(\ref{dWn}) and 
Eq.~(\ref{dWp}) 
are added, the first two terms of each expression will cancel each other.

To rewrite the terms involving the gravitational potentials in 
Eq.~(\ref{dWn}) and Eq.~(\ref{dWp}), we need 
to use 
\beq
\int n_\n \eta_\p^{i*} \nabla_i (\delta_{\xi_\p} \Phi_\p) dV = 
\int n_\p \xi_\p^i \nabla_i (\delta_{\eta_\p} \Phi_\n)^* dV \ .
\eeq
Then it is easy to show that the last two terms in Eq.~(\ref{dWp})
will cancel the last two terms in Eq.~(\ref{dWn}) once 
we add the two expressions.

In other words, we have proved that, in addition to (\ref{sfsymm1})--(\ref{sfsymm3}), the 
superfluid equations have the following symmetry
\beq
  \left<\eta_\n,D_\n \xi_\p\right> + \left<\eta_\p,D_\p \xi_\n\right>  =  \left< \xi_\n, D_\n 
\eta_\p\right>^* + \left<\xi_\p, D_\p \eta_\n \right>^* 
\eeq
and it follows that
\beq
W =  W_\n (\eta_\n,\xi_\n) + W_\p (\eta_\p,\xi_\p)
\eeq
is a conserved quantity.

By analogy with the single fluid case, it now makes sense to define the
canonical energy of the system as
\beq
E_c = { m_\B \over 2} \left[W_\n (\partial_t {\xi}_\n,\xi_\n) + 
W_\p(\partial_t {\xi_\p},\xi_\p) \right] \ .
\eeq
This can be written 
\begin{eqnarray}
E_c = { m_\B \over 2} \left\{ \left<\partial_t \xi_\n , A_\n 
\partial_t \xi_\n\right> 
+ \left<\partial_t \xi_\p , A_\p \partial_t \xi_\p\right> \right. \hspace{2cm} \nonumber \\
+  \left<\xi_\n, C_\n \xi_\n\right>^* + 
\left<\xi_\p,C_\p \xi_\p\right>^*  
+ \left. \left<\xi_\n,D_\n \xi_\p\right>^* + 
\left<\xi_\p,D_\p \xi_\n\right>^*  \right\} \ .
\label{ec2}
\end{eqnarray}
After some manipulations, we arrive at the following final 
explicit form
\begin{eqnarray}
E_c  = { 1 \over 2} \int \left\{ 
\rho_\n |\partial_t \xi_\n|^2 + \rho_\p |\partial_t \xi_\p|^2
- \rho_\n | v_\n^j \nabla_j \xi^\n_i|^2 \right. \hspace{3cm} \nonumber \\
- \rho_\p | v_\p^j \nabla_j 
\xi^\p_i|^2  + [\rho_\n \xi_\n^i \xi_\n^{j*} + \rho_\p \xi_\p^i \xi_\p^{j*}]\nabla_i 
\nabla_j \Phi
+ n_\n \xi_\n^i \xi_\n^{j*} \nabla_i \nabla_j \mu_\n \hspace{1cm}  \nonumber \\
+ 
n_\p \xi_\p^i \xi_\p^{j*} \nabla_i \nabla_j \mu_\p 
+ 
\left( {\partial {\mu}_\n \over \partial n_\n} \right)_{n_\p} 
|\delta n_\n|^2
+ \left( {\partial {\mu}_\p \over \partial n_\p} \right)_{n_\n} 
|\delta n_\p|^2 \hspace{1.5cm}  \nonumber \\
-  \left. { 1 \over 4\pi G} |\nabla_i \delta \Phi|^2 + 
\left( {\partial \mu_\n \over \partial n_\p} \right)_{n_\n} [ \delta 
n_\n \delta n_\p^*
+ \delta n_\n^* \delta n_\p]
\right\} dV \qquad  \qquad \ .
\label{sfcanerg}
\end{eqnarray}

It is worth noting that the symmetries of our system of equations imply the existence of a 
quadratic Lagrangian for the perturbations, which in turn implies the conservation of
$W=W_\n+W_\p$. We can derive the superfluid equations of motion (\ref{sfeom})  
from a variational principle whose action is (for clarity assuming that the 
displacement vectors are real, the generalisation to the complex case is straightforward)
\begin{eqnarray}
I = \int {\cal L} dV = { m_\B \over 2} \left[ \left< \dot{\xi}_\n,  A_\n \dot{\xi}_\n \right>  + 
\left< \dot{\xi}_\p, A_\p \dot{\xi}_\p \right>
+ \left< \dot{\xi}_\n, B_\n \xi_\n \right> \right. \nonumber \\
+  \left< \dot{\xi}_\p, B_\p \xi_\p \right>
- \left< \xi_\n, C_\n \xi_\n \right> - \left< \xi_\p, C_\p \xi_\p \right> \hspace{1cm} \nonumber \\
- \left< \xi_\p, D_\n \xi_\n \right> - \left< \xi_\n, D_\p \xi_\p \right>
\left. \right]  \ .
\end{eqnarray}
Then the momentum conjugate to $\dot{\xi}_\X$ follows from $\partial {\cal L} /\partial \dot{\xi}_\X$, 
while the equations of motion (\ref{sfeom}) can be derived from 
the standard Euler-Lagrange equations
\beq
\partial_t \left( { \partial {\cal L} \over \partial \dot{\xi}_\X } \right) -  { \partial {\cal L} \over \partial \xi_\X } = 0
\ .
\eeq 
Finally, the canonical energy for the system is
\beq
E_c = \int \left[  \dot{\xi}_\n^j { \partial {\cal L } \over \partial \dot{\xi}_\n^j} 
+ \dot{\xi}_\p^j { \partial {\cal L } \over \partial \dot{\xi}_\p^j}
- {\cal L } 
\right]  dV \ .
\eeq
One can readily verify that these formulas lead to the given results.

In an axisymmetric system
we can also define a conserved angular momentum;
\begin{eqnarray}
J_c = - { m_\B \over 2} W_\n (\partial_\varphi \xi_\n, \xi_\n) -  
{ m_\B \over 2} W_\p (\partial_\varphi \xi_\p, \xi_\p) \hspace{2cm} \nonumber \\
=  - m_\B \mbox{ Re } \left\{ \left< \partial_\varphi \xi_\n, A_\n 
\partial_t{\xi}_\n + { 1 \over 2} 
B_\n \xi_\n \right> \right. \hspace{1cm} \nonumber \\
+ \left. \left< \partial_\varphi \xi_\p, A_\p 
\partial_t{\xi}_\p + { 1 \over 2} 
B_\p \xi_\p \right>  \right\}  \ .
\label{jc2}
\end{eqnarray}

Now one can readily use Eq.~(\ref{ec2}) and Eq.~(\ref{jc2}) to prove 
that, for a normal mode solution to the problem, $\xi_\X = \hat{\xi}_\X 
e^{i(m\varphi+\omega t) }$, 
the canonical energy and angular momentum will still be related 
by Eq.~(\ref{EJrel}).  Furthermore, it follows that we must have 
$E_c=J_c=0$ for dynamically unstable  
(complex frequency) modes,  just like in the single fluid case. This is 
hardly surprising, but it could turn out to 
be a very useful result. One could, for 
example, hope to be able to use our expressions for the canonical energy
and angular momentum to derive necessary criteria for the superfluid 
two-stream instability \citep{acp03:_twostream_prl,prix04:_inertial,
andersson02:_twostream}.   

\subsection{Trivial displacements}

Before we close this section, let us address the issue of trivial 
displacements
in the two-fluid problem. We clearly need to deal with two sets of 
trivial displacements, one for each 
fluid degree of freedom. 
Fortunately, the analysis of these trivials is essentially identical
to that of the single barotropic fluid case discussed in 
Sec.~\ref{sinflu}, see \cite{friedman78:_lagran} for further details. 

In the superfluid problem, the trivial displacements are such that
\beq
\delta n_\X =  \delta v^\X_i = 0 \ .
\eeq 
It is easy to see from the superfluid perturbation equations that 
the two sets of equations that determine the functional form of the 
trivials are identical to the corresponding single fluid equations. 
This means that the single fluid result can be adapted to the superfluid 
problem: we simply have one trivial displacement ($\zeta_\X$) per fluid. 

In our case, we can take the requirement that the 
canonical displacements $\xi_\X$ must be 
``orthogonal'' to the trivials to mean that we should have
\beq
W_\n (\zeta_\n, \xi_\n)  = W_\p (\zeta_\p, \xi_\p) = 0 \ .
\eeq 
This condition leads to the trivial displacements having identical form 
to those of the single-fluid problem. However, it is worth noticing that 
we could in principle permit the somewhat less restrictive condition
\beq
W_\n (\zeta_\n, \xi_\n)  + W_\p (\zeta_\p, \xi_\p) = 0
\eeq
and still ensure that the trivial displacements do not affect our 
conserved quantities.  
We have not yet investigated the implications of this possibility.

Finally, and most importantly, 
one can readily extend the calculation of \cite{friedman78:_lagran} to 
prove that normal modes
are (usually) orthogonal to the trivials also in the present case. 
In other words, normal modes
may serve as canonical initial data.
This is extremely useful as we hope to use the canonical energy and 
angular momentum to assess the stability of various superfluid  normal mode 
solutions.

\section{Instabilities of rotating superfluid stars} \label{instab}

The main motivation for the present investigation was the lack of  
proper instability criteria for rotating superfluid stars. The
need for such criteria is clear given that it has long been acknowledged
that the astrophysical relevance of the gravitational-wave 
driven CFS instability (of both f- and r-modes) may 
depend on the extent to which superfluid dissipation [like mutual 
friction \citep{lindblom95:_does_gravit_radiat_limit_angul,lindblom00:_r_modes_superfl_neutr_stars}] 
counteracts the growth of the unstable mode. It would seem 
obvious that, before worrying about such issues, one ought to establish 
that the instability is actually present once the star becomes 
superfluid.  We are not aware of any such proof, despite the number of 
investigations of unstable oscillations of superfluid stars 
that exist in the literature. It should be clear that this is a 
non-trivial issue given the simple fact that the two fluids are only 
weakly coupled, and may in fact rotate at different rates. 

In this section we provide the first applications of our conserved
canonical energy and angular momentum for the two fluid problem. 
We first present an argument in favour of the simple single fluid 
criterion for CFS instability---that the instability sets in when the
pattern speed of an originally backwards moving 
mode passes through zero in the inertial frame---holding also in the 
superfluid problem. Having done this, we discuss the superfluid r-mode 
instability.   

\subsection{The superfluid CFS instability}

The main question here is: Does the simple criterion that a 
counter-rotating mode becomes unstable when the pattern speed changes 
sign remain valid also in the two-fluid problem?  Intuitively, one might 
expect this to be the case, but it nevertheless warrants a proof. 

We approach the problem as in the single fluid case. Assuming a real 
frequency mode solution to the perturbation equations, Eq.~(\ref{jc2}) 
leads to 
\begin{eqnarray}
J_c = -m \left\{ \omega [ \left<\xi_\n, \rho_\n \xi_\n\right> + 
\left<\xi_\p , \rho_\p \xi_\p\right>] \right.
\hspace{2cm} \nonumber \\
-\left.  [ \left<\xi_\n, i\rho_\n \vec{v}_\n\cdot\nabla \xi_\n\right> + 
\left<\xi_\p, i\rho_\p \vec{v}_\p\cdot\nabla \xi_\p\right>
]\right\} \ . 
\end{eqnarray}
Divide through to get 
\begin{eqnarray}
{ J_c \over \left<\xi_\n, \rho_\n \xi_\n\right> + \left<\xi_\p , \rho_\p 
\xi_\p\right> } \hspace{5.5cm} \nonumber \\  \qquad = 
-m\omega + m  {\left<\xi_\n, i\rho_\n \vec{v}_\n\cdot\nabla 
\xi_\n\right> + \left<\xi_\p, i\rho_\p \vec{v}_\p\cdot\nabla 
\xi_\p\right> \over \left<\xi_\n, \rho_\n \xi_\n\right> + \left<\xi_\p , 
\rho_\p \xi_\p\right> } \ .
\end{eqnarray}

Using cylindrical coordinates, we see from the results in 
Sec.~\ref{pfinstab} that
\beq
- i \rho_\n \xi_{\n i}^* v_\n^j \nabla_j \xi_\n^i = \rho_\n \Omega_\n 
[ m |\xi_\n|^2 + i (\xi_\n^*\times\xi_\n)_z] \ .
\eeq
We then know  that
\beq
 \rho_\n \Omega_\n (m-1) |\xi_\n|^2 \le - i \rho_\n \xi_{\n i}^* v^j 
\nabla_j \xi_\n^i
\le  \rho_\n \Omega_\n (m+1) |\xi_\n|^2
\eeq
and it is easy to see that we will have\footnote{We are assuming that 
the two fluids rotate in the same direction here. Although it is 
possible to construct models 
for counter-rotating backgrounds these cases are somewhat pathological, 
and we do not expect them to have any astrophysical relevance.}
\beq
 \left< \xi_\X, i\rho_\X \vec{v}_\X\cdot\nabla \xi_\X\right>  \ge - (m+1) 
\Omega_\X a_\X
\eeq
where we have defined $a_\X =  \left<\xi_\X, \rho_\X \xi_\X\right> > 0$.
We also get
\beq
\left< \xi_\X, i\rho_\X \vec{v}_\X\cdot\nabla \xi_\X\right> 
 \le - (m-1)  \Omega_\X a_\X \ .
\eeq

These results can be summarised as
\begin{eqnarray}
\sigma_p (a_\n + a_\p) - \left( 1 + {1\over m} \right) [ \Omega_\n a_\n 
+ \Omega_\p a_\p] 
\le  J_c/m^2    \hspace{1cm} \nonumber \\
\le
\sigma_p  (a_\n + a_\p) 
- {1\over 2} \left( 1 - {1\over m} \right)[ \Omega_\n a_\n + \Omega_\p 
a_\p] \ .
\end{eqnarray}
This is the relation  we need. Provided that 
$\Omega_\n a_\n + \Omega_\p a_\p >0$ (which is obviously 
true if the fluids both rotate in the positive direction)
we easily show that 
\begin{itemize}
\item if we let $[\Omega_n,\Omega_\p]\to 0$ while $\omega$ is finite, 
then for co-rotating
modes we have $\sigma_p>0$, which means that $J_c>0$. In contrast, for 
counter-rotating modes
$\sigma_p<0$ and $J_c<0$. In both cases, we will get $E_c>0$ which 
indicates that 
all (finite frequency) modes are stable.

\item if we consider a region near $\omega=0$ for finite rotation rates, 
then 
$\sigma_p=0$ implies that $J_c<0$. This means that the mode is stable 
as long as $\sigma_p<0$, but when $\sigma_p$ changes sign (and the mode 
becomes 
co-rotating) we will have $E_c<0$ and an instability.  
\end{itemize}
This concludes the proof that the criterion for the onset of
radiation driven instabilities of modes that have a finite frequency 
limit as $\Omega \to 0$  remains as in the single fluid case.
Modes become unstable when the (inertial frame) pattern speed changes 
sign. 

\section{The superfluid r-mode instability} \label{rinstab}

To conclude this paper we will discuss some aspects of the r-mode 
instability 
for rotating superfluid stars. Although we are not yet at a point
where we can discuss the general problem (since we did not 
include entrainment in our derivation), we can still learn quite a lot 
about the issues that arise when we consider two coupled fluids. 
Furthermore, we are not aware of any previous proof of the presence of 
an instability in the case when the two fluids rotate at different 
rates. 
The nature of the various inertial modes of oscillation (of which the 
r-modes
form a sub-class) of a superfluid star has, however, been discussed by 
several authors. We will draw on these investigations for information 
concerning the nature of the r-modes in both (i) the general case of 
a background star such that the two fluids rotate at different rates
(although with respect to the same axis), and (ii) the special case of  
co-rotating fluids.

\subsection{The case with relative rotation}

The general case, in which $\Omega_\n \neq \Omega_\p$, was recently 
discussed by \cite{prix04:_inertial}. From that study we learn that, in 
absence of entrainment, the r-mode fluid motion must be such that only 
one of the fluids oscillates. This means that 
we will have two classes of modes, corresponding to 
\begin{equation}
\delta \vec{v}_\X \neq 0 \ , \quad \delta \vec{v}_\Y = 0  \ , \quad 
\omega = m\Omega_\X \left[ 1 - { 2 \over l(l+1)} \right] \ .
\end{equation}

Since this implies that only one of the two displacements vectors is 
non-vanishing, it is straightforward to show that the canonical energy  
Eq.~(\ref{sfcanerg}) reduces to 
\begin{eqnarray}
E_c = { 1 \over 2} \int \left\{ 
\rho_\X |\partial_t \xi_\X|^2 
- \rho_\X | v^j \nabla_j \xi^\X_i|^2 \right. \hspace{1.5cm} \nonumber \\
\left. + \rho_\X\xi_\X^i \xi_\X^{j*}\nabla_i 
\nabla_j( \Phi
+ \tilde{\mu}_\X)
\right\} dV 
\end{eqnarray}
where we have assumed that $\delta n_\X$ and $\delta \Phi$ are of higher 
order in the slow-rotation scheme
(as in the single fluid case), and also used $v^i_\X=v^i$. 
Noticing the close resemblance of this result to the expression for the 
case of a barotropic single fluid, Eq.~(\ref{ec1}), we readily infer 
that both these classes of modes will be unstable due 
to the emission of gravitational radiation.  

However, this result is likely of very limited relevance. Even though it 
is expected that the two rotation rates will be slightly different in 
astrophysical neutron stars, the entrainment coupling will affect the 
r-modes significantly [as elucidated by \cite{prix04:_inertial}],  
naturally leading to both displacements being non-zero. This means that, 
despite being of conceptual interest, the case we have discussed here is 
pathological.

\subsection{The co-rotating case}

In the special case of $\Omega_\n=\Omega_\p= \Omega$, we know  
\citep{andersson01:_dyn_superfl_ns,comer02:_zero_freq_subspace,
lee03:_superfl_r_modes,
yoshida03:_r-modes_relat_superfl,prix04:_inertial} that there will exist 
two classes of r-modes. One is such that the two fluids move in phase, 
while the other has the two fluids counter-moving.  Furthermore, for 
non-stratified stars one can show that these two degrees 
of freedom decouple. Given the relative simplicity of this case, we will 
focus our attention on it. The general case essentially follows as a 
linear combination of the two results we present. 

We begin by introducing the two classes of displacements
\beq
 \xi_i^+ = {n_n \xi_i^\n + n_\p \xi_i^\p \over n_\n + n_\p}
\eeq
and
\beq
\xi_i^- = \xi_i^\n - \xi_i^\p \ . 
\eeq
These are clearly such that,  when $ \xi_i^\n = \xi_i^\p $ and the 
two fluids move together only $ \xi_i^+ $ is present,
while the total momentum flux vanishes when $\xi_i^+=0$. 

Now consider purely axial r-mode solutions such that $\xi_i^-=0$.  
In this situation the canonical energy Eq.~(\ref{sfcanerg})  
can be written as (since $v_\n^j = v_\p^j=v^j$)
\begin{eqnarray}
E_c \approx { 1 \over 2} \int  \left\{ \rho
 |\partial_t \xi^+|^2
-  \rho | v^j \nabla_j \xi^+_i|^2  +  \rho \xi_+^i \xi_+^{j*} \nabla_i \nabla_j 
\Phi \right.  \nonumber \\ 
 \left. +
\xi_+^i \xi_+^{j*} [n_\n  \nabla_i \nabla_j \mu_\n + 
n_\p  \nabla_i \nabla_j \mu_\p ]
\right\} dV 
\end{eqnarray}
where we (again) neglect the higher order contributions 
from $\delta n_\X$ and $\delta \Phi$. 

In this expression, the last term can be rewritten using the fact that 
we must have 
\begin{equation}
\nabla_i \mu_\n = \nabla_i \mu_\p \equiv \nabla_i \mu
\end{equation}
if the two fluids rotate at the same rate, cf Eq.~(\ref{eul1}).
This immediately leads to 
\begin{equation}
n_\n  \nabla_i \nabla_j \mu_\n + 
n_\p  \nabla_i \nabla_j \mu_\p =  n \nabla_i \nabla_j \mu
\end{equation}
(where $n = n_\n + n_\p$) and
\begin{eqnarray}
E_c = { 1 \over 2} \int \rho \left\{ 
 |\partial_t \xi^+|^2
-  | v^j \nabla_j \xi^+_i|^2 
+ \xi_+^i \xi_+^{j*} \nabla_i \nabla_j (\Phi+\tilde{\mu})
\right\} dV \ .
\end{eqnarray}
Clearly, this result is (provided that we identify $\vec{\xi}^+$ with 
$\vec{\xi}$)
identical to that of  the single fluid problem, Eq.~(\ref{ec1}). 
Hence, the instability of pure $\vec{\xi}^+$ modes follows from the 
calculation in Section~3.4. This is not very surprising given that the 
degree of 
freedom we are considering is such that the two fluids move together.

Next we consider the canonical energy for counter-rotating modes, which 
are such that
\beq
n_\n \xi_\n^i + n_\p \xi_\p^i = 0 \ .
\eeq
In this situation only $ \xi_i^- $ is present,  and we have
\begin{eqnarray}
\xi_i^{\n} &=& {n_\p \over n}\xi_i^- \ , \\
\xi_i^{\p} &=& - {n_\n \over n_\p} \xi_i^{\n} = - {n_\n \over n} \xi_i^-
\ .
\end{eqnarray}
Incidentally, the latter of these relations emphasizes the fact that 
pure $\vec{\xi}^-$ modes can only exist for non-stratified stars. 
We know from the results of \cite{prix04:_inertial} that $\xi_i^{\n}$ 
and $\xi_i^{\p}$
will have the same functional dependence on the radial coordinate $r$.
This means that, in order for both displacements to be non-zero, they 
must be 
proportional. This is clearly only possible if $n_\n/ n_\p$ is constant.

Using the above relations in the expression for the canonical energy 
Eq.~(\ref{sfcanerg}), 
one can show that 
\begin{eqnarray}
E_c = { 1 \over 2} \int \rho_\n x_\p \left\{ 
  |\partial_t \xi_i^-|^2
+ | v^j \nabla_j \xi_i^-|^2  \hspace{1cm} \right. \nonumber \\ 
\left. +  \xi^i_- \xi^{j*}_- \nabla_i \nabla_j ( \Phi + \tilde{\mu})
\right\} dV 
\end{eqnarray}
where $x_p = {n_\p \over n}$ is the proton fraction.
Since the expression in the bracket has the same form 
as in the single fluid case, and the prefactor $\rho_\n x_\p$ is 
positive definite, it is easy to prove that 
$ E_c < 0 $ also for these counter-moving modes.

\subsection{Gravitational-wave emission} 

At this point it is appropriate to discuss how efficient the
oscillations of a superfluid star are as a source of gravitational 
waves. After all, we have shown that the superfluid r-modes 
generally lead to $E_c < 0$, and should therefore be driven unstable by 
radiation.
The relevance of the instability is then largely dependent 
on the rate at which the motion generates gravitational
radiation. 

To address this problem, we consider a source with weak internal 
gravity, and focus our attention on a single pulsation mode with 
time-dependence $\exp(i\omega t)$.  We also assume 
that the background is such that the two fluids rotate at different 
rates: $\Omega_n$ and $\Omega_p$, respectively.
  
The gravitational-wave luminosity follows from e.g.
\cite{1980RvMP...52..299T}, 
\begin{equation}
    {dE \over dt} = \sum_{l=2}^\infty N_l \omega^{2 l + 2} 
    \left( | \delta D_{lm} |^2 + | \delta J_{lm} |^2 \right) \ ,
    \label{gwlum}
\end{equation}
where 
\begin{equation}
     N_l = {4\pi G \over c^{2 l + 1} } { (l + 1) (l + 2) \over l 
           (l - 1)[(2 l + 1)!!]^2 } \ .
\end{equation}
The first term in the bracket of Eq.~(\ref{gwlum}) represents radiation 
due to the mass multipoles.  
These are, quite generally, determined from
\begin{equation}
   \delta D_{lm} = \int T_{00} Y_{lm}^* r^l dV 
\end{equation}
where $T_{\mu \nu}$ is the contribution to stress-energy tensor 
associated with non-axisymmetric motion in the source.
The second term 
in the bracket of Eq.~(\ref{gwlum}) corresponds to the current 
multipoles, which follow from 
\begin{equation}
   \delta J_{lm} = \int (- T_{0j} )  Y_{j,lm}^{B*} dV \ .
\end{equation}
where $\vec{Y}_{lm}^{B}\propto \hat{r} \times \nabla Y_l^m$ are the 
magnetic multipoles \citep{1980RvMP...52..299T}.

By taking the Newtonian limit of the 
relativistic stress-energy tensor 
(see, for example, \cite{comer02:_zero_freq_subspace})
\begin{equation}
    T_\mu^\nu = \Psi \delta_\mu^\nu + p^\nu \chi_\mu + n^\nu \mu_\nu
\end{equation}
[which is easily done using formulas given in Appendix~A of 
\cite{andersson01:_dyn_superfl_ns}], one can show that 
\begin{equation}
   T_{00} \approx (\rho_n + \rho_p)   
\end{equation}
and
\begin{equation}
   T_{0j} \approx  \rho_n \vec{v}_n + \rho_p 
   \vec{v}_p 
\end{equation}
for a Newtonian source. Perturbing these expressions we find that
\begin{equation}
     \delta D_{lm} = \int (\delta \rho_\n + \delta \rho_\p) r^l 
                     Y_{lm}^* dV \label{mass}
\end{equation}
and
\begin{eqnarray}
    \delta J_{lm} = {2 \over c} \sqrt{ {l \over l + 1}} \int r^l 
    (\rho_\n \delta \vec{v}_\n + \rho_\p \delta \vec{v}_\p \hspace{1cm} \nonumber \\
+ \delta 
    \rho_\n \vec{\Omega}_\n + \delta \rho_\p \vec{\Omega}_\p) \cdot 
    \vec{Y}_{lm}^{B*} dV \ . \label{curr}
\end{eqnarray} 

As discussed in the previous section, it is sometimes instructive to 
express the superfluid formulas in terms of the variables
\begin{equation}
    \delta \vec{v}^+ = { \rho_\n \over \rho} \delta \vec{v}_\n + 
    { \rho_\p \over \rho} \delta \vec{v}_\p 
\end{equation}
and
\begin{equation}
    \delta \vec{v}^- = \delta\vec{v}_\p - \delta\vec{v}_\n \ .
\end{equation}
Using these, together with
\begin{equation}
    \delta \rho = \delta \rho_\p + \delta \rho_\n
\end{equation}
we get 
\begin{equation}
    \delta D_{lm} = \int \delta \rho  r^l Y_{lm}^* dV \ .
\end{equation}
and
\begin{eqnarray}
    \delta J_{lm} = {2 \over c} \sqrt{ {l\over l + 1}} \int r^l 
    [ \rho \delta \vec{v}^+ + \delta \rho \vec{\Omega_\p} \hspace{1.5cm} \nonumber \\
 + \delta 
    \rho_n (\vec{\Omega}_\n - \vec{\Omega}_\p)] \cdot 
    \vec{Y}_{lm}^{B*} dV \ . \label{curr2}
\end{eqnarray}
When written in this form, the formulas closely resemble the standard
single-fluid results. The only new feature is the last term in the 
bracket of Eq.~(\ref{curr2}).  

An interesting question concerns whether 
it is possible to have oscillations in a superfluid star that (at 
least at this post-Newtonian level) do not radiate gravitationally. 
From the above equations we immediately deduce that 
in the case of a non-rotating star (with $\Omega_\n = 
\Omega_\p=0$) or a co-rotating star  (when $\Omega_\n = 
\Omega_\p$) we must have $\delta \rho = \delta \vec{v}^+ = 0$ 
in order not to have any gravitational radiation emission. 
This result is quite intuitive since, as discussed in the previous 
section, 
the co-moving degree of freedom represents the total momentum flux.
It is not  surprising to find that motion which corresponds to zero
momentum flux does not radiate gravitationally.  

However, when combined with the results obtained from our superfluid 
canonical 
energy, these results illustrate some of the complexities associated 
with a discussion of gravitational-wave driven instabilities in a 
superfluid star. In particular, we conclude that even though they are 
formally unstable ($E_c<0$) any purely  counter-moving modes (for which 
$\vec{\xi}^+=0$) will not grow ($dE/dt=0$). 


\section{Concluding remarks} \label{conclude}

With this paper we have taken the first steps towards a
Lagrangian perturbation framework for rotating 
non-relativistic superfluids. The primary motivation for this 
work, and the key application concerns the stability properties of rotating superfluid 
neutron stars. Our analysis generalises the classic work of 
\cite{friedman78:_lagran,friedman78:_secul_instab} to the case of stars which 
require a multi-fluid description.  
We have applied our framework to the problem 
of dynamical and secular instabilities of a simplified superfluid 
neutron star model in which the two fluids are allowed to have different 
uniform rotation rates around the same axis, and where the entrainment effect is neglected.  
We have demonstrated that 
the criterion for the onset of radiation driven instabilities for modes 
that have a finite frequency as the background rotation vanishes 
remains unchanged from the ordinary fluid case.  We have also considered 
the superfluid analogue of the r-mode instability, and found that 
both co- and counter-moving superfluid r-modes have negative canonical energy and 
will therefore be driven unstable by gravitational-wave emission. 

Of course, our neglect of the entrainment is a serious limitation of 
the formalism.  However, as we stated earlier, if we could not make 
progress in the case of vanishing entrainment it would be pointless to 
proceed to the general case.  Naturally, we are encouraged by the progress 
we have made, which provides crucial benchmarks for the onset 
of instability in multi-fluid systems.  In the near future, 
we will aim to generalize our results to include entrainment.  This 
is absolutely essential for the study of gravitational-wave driven 
instabilities, since the primary superfluid damping mechanism is the 
mutual friction which depends crucially on the entrainment.  The inclusion 
of entrainment will also be important for future studies of the 
superfluid two-stream instability, since entrainment may be the dominant 
coupling to drive the instability in the interiors of neutron stars.

\section*{Acknowledgments}

We thank John Friedman for reading a draft
version of the manuscript and providing useful comments that improved 
the presentation. We also thank Reinhard Prix for useful 
discussions of the two-fluid model for superfluid neutron stars.  NA 
gratefully acknowledges support from PPARC grant PPA/G/S/2002/00038 
and the Leverhulme
trust via a prize fellowship.  GLC gratefully acknowledges 
support from NSF grant PHYS-0140138.

\end{document}